\documentclass[12pt,preprint]{aastex}

\newcommand\refp[1]{(\ref{#1})}

\shorttitle{Generalized Lagrangian points} 

\begin{document}

\title{A generalization of the Lagrangian points: \\ studies of
resonance for highly eccentric orbits}
 
\author{Margaret Pan and Re'em Sari}

\affil{130-33 Caltech, Pasadena, CA 91125}
\begin{abstract}  
We develop a framework based on energy kicks for the evolution of
high-eccentricity long-period orbits with Jacobi constant close to 3
in the restricted circular planar three-body problem where the
secondary and primary masses have mass ratio $\mu\ll 1$. We use this
framework to explore mean-motion resonances between the test particle
and the secondary mass. This approach leads to (i) a redefinition of
resonance orders to reflect the importance of interactions at
periapse; (ii) a pendulum-like equation describing the librations of
resonance orbits; (iii) an analogy between these new fixed points and
the Lagrangian points as well as between librations around the fixed
points and the well known tadpole and horseshoe orbits; (iv) a
condition $a\sim \mu^{-2/5}$ for the onset of chaos at large semimajor
axis $a$; (v) the existence of a range $\mu<\; \sim\! 5\times 10^{-6}$
in secondary mass for which a test particle initially close to the
secondary cannot escape from the system, at least in the planar
problem; (vi) a simple explanation for the presence of asymmetric
librations in exterior $1:N$ resonances.
\end{abstract}

%%%%%%%%%%%%%%%%%        1.  INTRODUCTION          %%%%%%%%%%%%%%%%%
\section{Introduction}

The three-body problem, or the dynamics of three masses due to their
mutual gravitational influences, has a number of well-known special
cases. One of these, the circular planar restricted case, requires
that the primary and secondary bodies, $m_1$ and $m_2$, follow
circular orbits about their common center of mass and that the third
body be a massless test particle moving in the massive bodies' orbit
plane. These conditions simplify the three-body problem enough to
produce an integral of the motion: the Jacobi constant
$C_J=-2(E-\Omega h)$ where $E$ is the particle's energy%
\footnote{We refer to the test particle as `the particle' and to its
energy per unit mass and angular momentum per unit mass as its
`energy' and `angular momentum'.},
$h$ is the particle's angular momentum, and $\Omega$ is the massive
bodies' constant angular velocity.

Still, the circular planar restricted case has important applications
to the dynamics of our solar system. Many of the orbits of major
planets about the sun are nearly circular and are roughly confined to
a plane; the same goes for many of the orbits of large moons about
their planets. Common examples of applications for the circular planar
restricted case include the effects of Jupiter on the asteroid belt;
of Neptune on the Kuiper belt; of moons on planetary rings; and of
giant planets on comets. 

This paper describes a study of this problem in the regime where $m_2
\ll m_1$, the test particle's eccentricity is large, and the Jacobi
constant is greater than but close to $3$ in the standard system of
units where $G=1$, the primary-secondary separation is 1, $1 =
m_1+m_2\simeq m_1$, and, therefore, $\Omega=1$. Since values of $C_J$
near 3 correspond to test particles on circular orbits close to the
secondary, this special regime includes particles originally in
circular orbits around a star close enough to a planet for the planet
to perturb them into very eccentric orbits. Our interest in this
regime arises from an intent to investigate the paths through which
small particles are perturbed by a planet until they escape or are
captured. This problem was studied by \citet{ebf01} and \citet{far96}
via numerical simulations of three massive bodies in three
dimensions.
%Applications of the results shown here to the escape/capture problem 
%will be reported elsewhere, but 
Due to this motivation we use `star' and `planet' to refer to the
primary and secondary in the remainder of this paper.

In \S\ref{DELTAE1} we derive to first order in $\mu=m_2/(m_1+m_2)=m_2$
the energy kick received by a particle in a highly eccentric orbit
with semimajor axis $a\gg 1$ at each periapse passage. We show that
since the interaction is localized at periapse, this energy kick is
essentially independent of $a$ and depends only on the periapse
distance and the azimuth difference between the planet and particle at
periapse. In \S\ref{RESONANCES}, \S\ref{mapping}, and
\S\ref{higher-order} we use these energy kicks to find `fixed'
particle orbits and describe motion near them. These `fixed' orbits
are located at planet-particle mean-motion resonances. When observed
stroboscopically at periapse only, they appear as fixed points just
like the well-known Lagrangian points. We use both a continuous
approximation and a discrete mapping in a derivation of the particle's
motion around resonances, the resonance widths, and the libration
periods. When these librations are observed stroboscopically, they
likewise become analogies of the well-known tadpole and horseshoe
orbits. In \S\ref{CHAOS} we discuss types of chaos for
large-eccentricity orbits, and in \S\ref{DISCUSSION} we summarize and
discuss our findings.

%%%%%%%%%%%%%%%%%      2. 1ST-ORDER ENERGY KICK      %%%%%%%%%%%%%%%%%
\section{\label{DELTAE1} Energy kick to first order in {\boldmath $\mu$}}

Let $\Delta E$ the changes in the particle's energy between its
consecutive apoapse passages. In our units, where the angular velocity
of the star-planet is set to unity, the change in angular momentum is
also $\Delta E$ 
\footnote{Since the angular momentum is always perpendicular to the
orbit plane, only one of its components is nonzero. We therefore treat
the torque and the change in angular momentum as scalars equal to the
components of the vector torque and the vector change in angular 
momentum which are perpendicular to the orbit plane.}.
Therefore, it can be calculated by integrating the torque exerted on
the particle:
\begin{equation}
\Delta E = \int -\left. {\partial V\over \partial f} \right|_{r} dt
\end{equation}
where $V$ is the gravitational potential produces by the planet and
star and $f$ is the particle's azimuth in inertial space.

To begin with, we estimate the energy kicks to first order in
$\mu$. We express $\Delta E$ as $\Delta E=\mu\Delta E^1 +
\mathcal{O}(\mu^2)$. To evaluate $\Delta E^1$ we calculate the torque
assuming that the particle moves on a Keplerian trajectory around the
star, with its focus at the center of mass. The effect of the
deviation of the trajectory from that description on $\Delta E$ is of
order $\mu^2$ or higher.

Since we are considering only the time elapsed between two consecutive
apoapses, we choose coordinates such that the time $t=0$ when the
particle is at periapse and the direction of periapse is along the
positive $x$-axis. The planet and star are in uniform circular motion,
so we can write $V=V(\theta,r)$ where $\theta$ is the angle between
the planet and the particle and $r$ is the particle's distance from
the origin. This gives
\begin{eqnarray}
\label{deltaE}
\Delta E = \int \left. {\partial V\over \partial \theta}\right |_r\; dt 
\;\;\; .
\end{eqnarray}
$V$ is given explicitly by
\begin{equation}
V = V_{\rm planet} + V_{\rm star}
  = {\mu \over |\vec{r}-\vec{r}_{\rm planet}|} 
      + {1\over |\vec{r}-\vec{r}_{\rm star}|} \;\;\; ;
\end{equation}
to first order in $\mu$, this gives
\begin{equation}
\label{V1}
V = {1\over r} + \mu \left ({1 \over (r^2 + 1 - 2r\cos\theta)^{1/2}}
                             - {\cos\theta\over r^2} \right ) \;\;\; .
\end{equation}

Let $\phi$ be the angle between the planet and the particle at
periapse%
\footnote{Thus defined, $\phi$ is the usual resonant argument measured
at periapse only.}, 
so that $\theta = \phi+t-f$. Then the derivative with respect to
$\theta$ at fixed $r$ which appears in Eq.~\refp{deltaE} can be
replaced by a derivative with respect to $\phi$. To first order in
$\mu$ the particle trajectory $r(t)$ can be assumed fixed and
independent of $\phi$, so we can move the $\phi$ derivative outside
the integral of Eq.~\refp{deltaE}. Using the first order expression
for $V$ we get
\begin{equation}
\Delta E^1 = -{dU^1\over d\phi}
\end{equation}
where the effective potential $U^1$ is given by
\begin{equation}
U^1 = -\int \left [{1\over \sqrt{r^2 + 1 - 2r\cos \theta}}
                     - {1\over \sqrt{r^2 + 1 + 2r\cos (t-f)}}
                     + {\cos \theta - \cos (t-f) \over r^2}
              \right ]\; dt \;\;\; .
\label{deltah}
\end{equation}
The integral is performed over one Keplerian orbit of the
particle. 

In this expression for $U^1$, the first term in the brackets is the
`direct' term; it represents the planet's contribution. The second
term does not contribute to $\Delta E^1$; it keeps $U^1$ from
diverging when $a \rightarrow \infty$ and is obtained from the first
term by substituting $\phi=\pi$. The third term is the `indirect'
term; it represents interactions with the star. $\Delta E^1$ and its
effective potential $U^1$ are functions of $\mu$, $\phi$, and the
particle trajectory shape, which determines $r$ and $f$ as a function
of $t$. Note that up to a constant, the effective potential $U^1$ is
simply the time-integrated potential over the trajectory of the
particle.

%%%%%%%%%%%%%%%%%  2.b. Large a (parabolas/escapes) %%%%%%%%%%%%%%%%%
%\subsection{The large-{\boldmath $a$} limit}

When the apoapse distance $a(1+e)$ is much larger than both 1 and the
periapse distance $r_p=a(1-e)$, the perturbing effects of the star and
planet on the particle near periapse dominate over perturbing effects
on the particle elsewhere in its orbit. In this regime, the entire
energy kick $\Delta E$ occurring between consecutive apoapse passages
can be thought of as a discrete event associated with a particular
periapse passage. In the limit as $a$ diverges due to energy kicks but
$C_J$ remains constant, $e\rightarrow 1$, $r_p$ approaches a constant,
and except near apoapse the entire trajectory approaches a parabola
independent of $a$: $a\rightarrow\infty$. If the particle is outside
the planet's Hill sphere,
\begin{equation}
\lim_{a\rightarrow\infty} r_p
  = \lim_{a\rightarrow \infty}
         \left [-{1\over 2}\left (C_J+{1\over a}\right )
                 {1\over \sqrt{1+e}}\right ]^2
  = {C_J^{\;2}\over 8}
\end{equation}
\begin{equation}
r=2r_p/(1+\cos f)
\end{equation}
\begin{equation}
{df \over dt} = {(1+\cos f)^2 \over (2r_p)^{3/2}}
\label{fdot}
\end{equation}
\begin{equation}
t = (2r_p)^{3/2}\cdot {1\over 6}\tan {f\over 2} 
                      \left (3+\tan^2 {f\over 2}\right ) \;\;\; .
\label{t(f)}
\end{equation}
For particles that start close to the orbit of the planet, the periapse
distance is therefore $r_p=9/8$.

\begin{figure}[!ht]
\begin{center}
\epsscale{.75}
\plotone{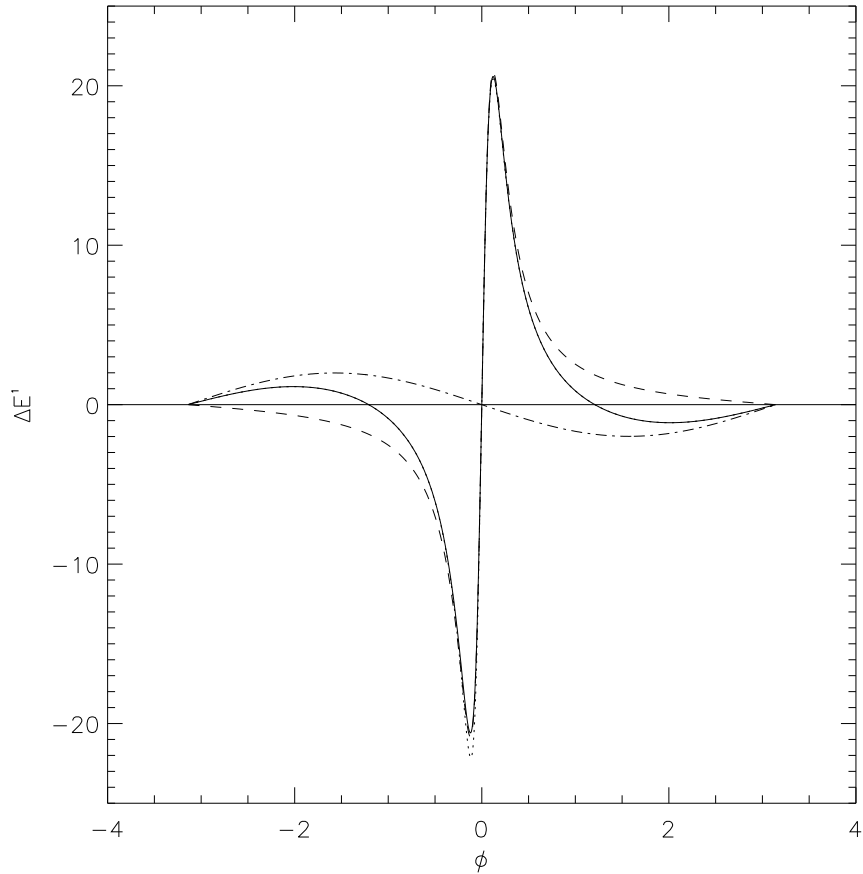}
\caption{First-order energy kick $\Delta E^1$ in the large-$a$ regime
(solid line) with $C_J=3$. The dotted line is the energy kick $\Delta
E/\mu$ calculated for a $\mu=10^{-3}$, $C_J=3$ parabolic orbit with
all higher order terms included. For this $\mu$, the first-order term
clearly dominates for all values of $\phi$; higher-order effects in
$\mu$ are visible only near $\phi = -0.12$. The dashed line is the
planet's direct contribution to $\Delta E^1$; the dash-dotted line is
the indirect contribution to $\Delta E^1$ from the star's reflex
motion.}
\label{deltaE_mu001}
\end{center}
\end{figure}

\begin{figure}[!ht]
\begin{center}
\epsscale{.75}
\plotone{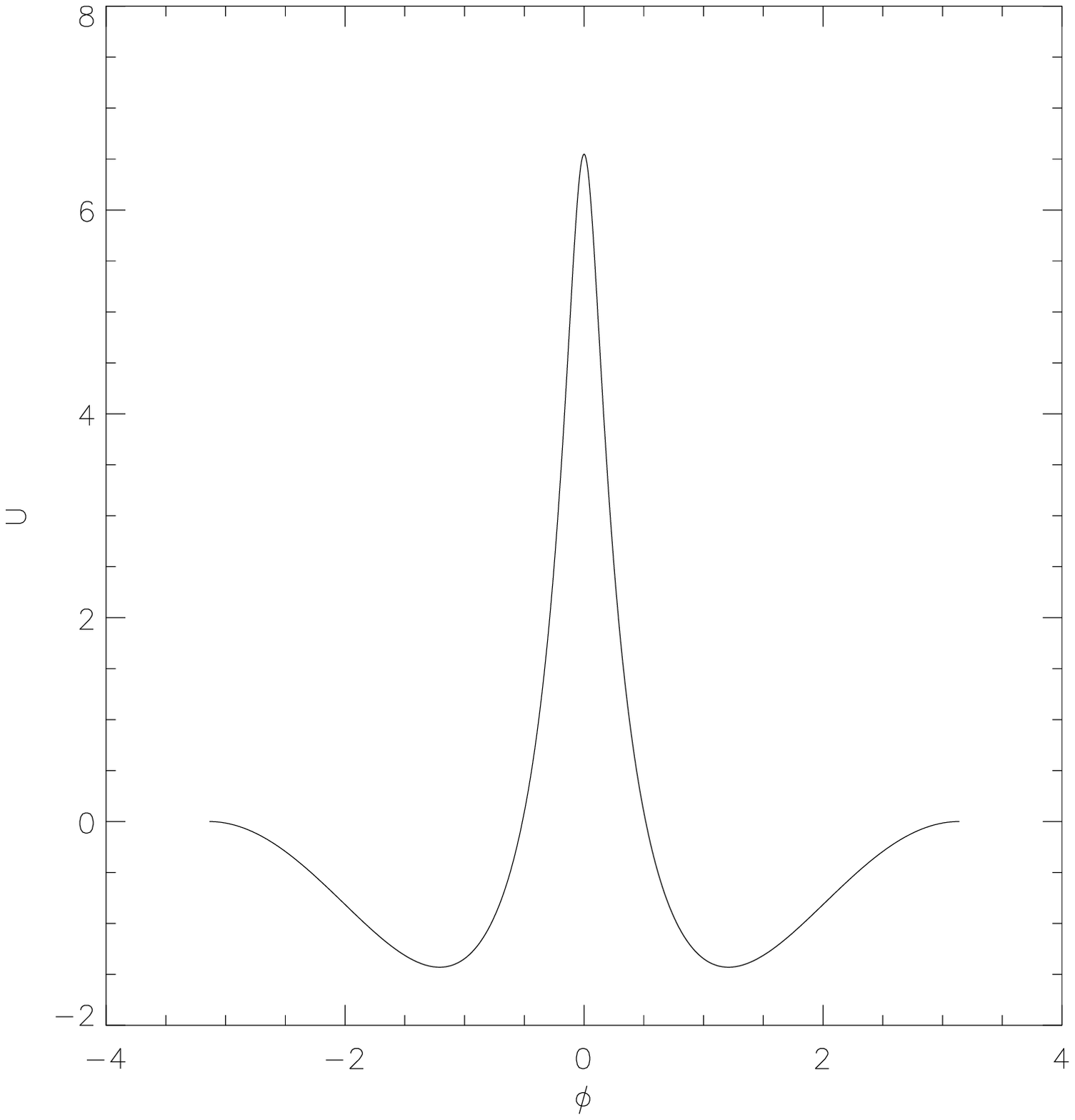}
\caption{Potential $U^1$ in the large-$a$ regime with $C_J=3$.}
\label{U_mu001}
\end{center}
\end{figure}

Given this asymptotic form for the orbit, we can calculate the
asymptotic forms of $U^1(a,e,\phi) \rightarrow U^1(r_p,\phi)$ and
$\Delta E^1(a,e,\phi) \rightarrow \Delta E^1(r_p,\phi)$ in the
large-$a$ limit.  For $C_J=3$, the computed values of $U^1$ and its
derivative $\Delta E^1$ as a function of $\phi$ are shown in
Figures~\ref{deltaE_mu001} and \ref{U_mu001}.  Near $\phi=0$, $\Delta
E^1$ is dominated by the direct particle-planet interaction because
the minimum planet-particle distance is much less than the
star-particle distance. When $\phi=0$, $\Delta E^1=0$ because of
symmetry.  When $\phi < 0$ but $|\phi|\ll 1$, the planet lags the
particle for most of the time the particle spends near periapse, so
$\Delta E^1 <0$. Similarly, when $\phi > 0$ and $|\phi|\ll 1$, $\Delta
E^1 >0$.

When $||\phi| - \pi|\ll 1$, the indirect contribution $\Delta E^1_{\rm
ind}$ due to the star's reflex motion dominates because the star
passes closer to the particle. From the $\phi$-dependent part of
star's contribution to the integral in Eq.~\refp{deltah}, $\Delta
E_{\rm ind}^1$ is a sinusoidal function of $\phi$:
\begin{equation}
\label{deltaE_star}
{\partial\over\partial\phi} 
      \int_{-\infty}^\infty {\cos\theta\over r^2}\; dt %&
  = -{\sin\phi\over\sqrt{2r_p}} \int_{-\pi}^\pi \cos(t-f)\; df 
  \simeq -2.0\sin\phi
\end{equation}
where in the last step we use $r_p=9/8$ as an example in evaluating
the coefficient.

The integral in Eq.~\refp{deltaE_star} seems to suggest that
star-particle interactions over all intervals in $f$ should contribute
significantly to $\Delta E^1_{\rm ind}$. However, as
Eq.~\refp{t(f)} shows, $|t|$ increases much faster than $|f|$ as $|f|$
approaches $\pi$. As a result, oscillations in $\cos(t-f)$ kill
contributions to the integral at $|f|$ near $\pi$, and the
star-particle interaction is important only near periapse.

We can also get the total contribution of terms that are second order
or above in $\mu$: this is just the difference between values of
$\Delta E$ found by numerical integration of the equations of motion
and values of $\mu\Delta E^1$ given by Eq.~\refp{deltah} (see
Figure~\ref{deltaE_mu001}). 

%We denote this correction to $\mu\Delta E^1$ as $\mu^2\Delta E^{\rm
%higher} = \Delta E - \mu\Delta E^1$ and discuss its effect on the
%particle's behavior further in \S\ref{higher-order}.

%%%%%%%%%%%%%%%%%          3. RESONANCES          %%%%%%%%%%%%%%%%%
\section{\label{RESONANCES}First-order resonances}

Resonances occur when the particle completes $p$ orbits in exactly the
time needed for the planet to complete $p+q$ orbits for some integers
$p$, $q$. This situation is known as a $p:p+q$ resonance. In the
standard treatment of these resonances, both orbits in question are
usually nearly circular and a significant interaction occurs every
time the bodies are at conjunction, i.e.~whenever their azimuths
coincide. This happens once every resonant cycle if $q=1$, so $q=1$
resonances are usually termed `first order' resonances. During a
conjunction between a test particle and planet in orbits with low
eccentricity $e \ll 1$, the torque exerted on the particle while the
particle precedes the planet almost cancels the torque exerted while
the particle lags the planet; the residual is of order $e$. When
$q>1$, $q$ conjunctions occur during each resonant cycle. Because they
occur in different positions in inertial space, their effects tend to
cancel each other, leaving a residual torque of order $e^q$.  Since
the interaction strength decreases exponentially with increasing $q$
as $e^{q}$, resonances in the standard treatment are usually
classified by $q$ value. Accordingly, a $p:p+q$ resonance is called a
`$q$th-order' resonance regardless of the value of $p$.

However, the high eccentricities of orbits in the large-$a$ regime
discussed here make the standard definition of resonance order
meaningless. Since $e\rightarrow 1$, resonances of different order
under the standard definition have comparable significance because
$e^q \simeq 1$. Also, encounters at periapse are physically far more
important than conjunctions at other points in the particle's
orbit. We therefore redefine the `orders of resonance' to focus on
interactions at periapse. If the planet completes an integer number of
orbits in the time it takes the particle to orbit exactly once, then
we say the particle is in a `first-order' resonance. In general, we
say the particle is in an `$p$th-order' resonance if the planet
completes an integer number of orbits in the time it takes the
particle to orbit $p$ times: then there are $p$ interactions within
one resonant cycle. In terms of the standard resonance treatment, we
say a $p:p+q$ resonance in the large-$a$ regime is `$p$th-order'
regardless of the value of $q$. In both the large- and
small-eccentricity cases, the order of the resonance is given by the
number of significant interactions within a single resonant cycle.

In the following we show that this revised definition does indeed make
sense. We calculate the widths of resonances of various orders in the
large-$a$ limit and show that with this new definition, their widths
decay exponentially with the order of the resonance. We discuss in
detail the first-order or $1:N$ resonances.

According to this definition of resonance orders, $\phi$ should be
constant in time if we consider a particle exactly at a first-order
resonance of semimajor axis $a_{\rm res}$ and if we ignore the
effects of energy kicks. A particle close to resonance with, say,
semimajor axis $a=a_{\rm res}+\Delta a$ should drift in $\phi$
over time at a constant rate, again ignoring energy kicks. The amount
of drift per orbit of the test particle is just the difference between
its orbital period $2\pi a^{3/2}$ and the resonant one $2\pi
a_{\rm res}^{3/2}$. We can express this drift as
\begin{equation}
{d\phi\over dn} = 3\pi a^{1/2}\Delta a \;\;\; .
\label{dtheta0dn}
\end{equation}
Here the $dn$ acts as a place-holder indicating that the $d\phi$ is
associated with a single particle orbit. The differential is justified
assuming $\Delta a \ll a^{-1/2}$ so that the change in $\phi$ per
particle orbit is much less than $\pi$. We refer to this differential
form as the continuous approximation.

Energy kicks cause the semimajor axis to evolve in time. To first
order in $\mu$ we have
\begin{equation}
\label{ddeltaadn}
{d(\Delta a)\over dn} = 2a^2{dE\over dn} =- 2a^2\mu {dU^1 \over d\phi}
%\Delta E^1 
\;\;\; .
\end{equation}
To justify the differentials here we require that $\mu$ be small
enough for the change in $\Delta a$ due to a single kick to be much
less than the typical $\Delta a$. We differentiate
Eq.~\refp{dtheta0dn} and substitute Eq.~\refp{ddeltaadn} to get
%\begin{eqnarray}
\begin{equation}
\label{d2theta0dn2}
{d^2 \phi \over dn^2} %& 
%   \simeq&
%   \simeq 3\pi\sqrt{a}{d (\Delta a)\over dn}
%           + \Delta a{3\pi\over 2}{1\over \sqrt{a}}{da\over dn}\;\;
%   \simeq \;\; 3\pi\sqrt{a}\cdot \mu\Delta E^1 \cdot 2a^2 \\
%&       =& 
   = -6\pi a^{5/2}\mu {dU^1 \over d\phi} 
%\Delta E^1 
\;\;\; .
\end{equation}
%\end{eqnarray}
This shows that $\phi$ simply evolves as a particle moving in the
potential $U^1(\phi)$. 

\subsection{Generalized Lagrangian points}

Since $U^1$ has four extrema at the four zeroes of $\Delta E^1$, there
are four fixed points. These fixed points are simply periodic orbits
of the particle. Two of the fixed points are unstable since they
correspond to maxima of the potential $\theta=0,\pi$. The other two
are stable since they correspond to potential minima at $\theta=\pm
1.21$.  The existence of two extrema at $\theta=0,\pi$ is guaranteed
by symmetry arguments.  The two additional extrema at $\phi=\pm 1.21$
occur where the energy kicks from the planet and star cancel each
other exactly. These extrema therefore appear only when the indirect
term---or, equivalently, the star's motion---is taken into account.

This discussion suggests an analogy between the five well known
Lagrangian points and the new fixed points. The two stable points
correspond to the stable Lagrangian points $L_4$ and $L_5$, which also
appear only when the motion of the star, i.e.~the indirect term, is
taken into account. The unstable fixed point at $\phi=\pi$ is the
analogue of $L_3$; the one at $\phi=0$ corresponds to $L_1$ and $L_2$,
which merge in this generalization. For a given resonance $1:N$,
$N=a^{3/2}$, we therefore denote the fixed points by $L_{12}^N$,
$L_3^N$, $L_4^N$, and $L_5^N$. The positions of these new fixed points
in comparison to their standard Lagrangian counterparts is summarized
in Table~\ref{reswidths}. 

\begin{table}[!ht]
\begin{center}
\begin{small}
\begin{tabular}{c||c||c|c|c|c}
                        &  Lagrangian points                  & \multicolumn{4}{c}{generalized Lagrangian points}       \\ 
\hline
resonant index          &      $(N=1)$                        &  $N=2$        &  $N=3$   &   $N=4$    & large $a$    \\ 
\hline\hline
semimajor axis          &        1                            & $2^{2/3}$    &  $3^{2/3}$& $4^{2/3}$  &  $a=N^{2/3}$    \\ 
\hline
physical meaning        &  particle is {\it stationary}       & \multicolumn{4}{c}{particle moves on {\it periodic orbit}}       \\
of fixed points         &  in rotating frame                  & \multicolumn{4}{c}{in rotating frame}                 \\ 
\hline
definition of           &  azimuth of particle                & \multicolumn{4}{c}{azimuth of particle in rotating frame} \\
angular variable        &  in the rotating frame              & \multicolumn{4}{c}{when it is at periapse}               \\ 
\hline
 $L_1$ \& $L_2$         &  $\phi_1=0$, $\phi_2=0$             & \multicolumn{4}{c}{single point $L_{12}^N$ with $\phi_{12}=0$}\\
 $L_3$                  &  $\phi_3=\pi$                       & \multicolumn{4}{c}{$\phi_3=\pi$} \\
 $L_4$ \& $L_5$ ($\phi_5=-\phi_4$) &$\phi_4=\pi/3\simeq 1.04$ & $\phi_4^2=1.196$ &  $\phi_4^3=1.196$ & $\phi_4^4=1.198$ & $\phi_4=1.21$ \\ 
\hline
min. tadpole period     & ${4\pi\over 3\sqrt{3}}\mu^{-1/2}\simeq 2.42\mu^{-1/2}$& $4.4\mu^{-1/2}$ & $5.1\mu^{-1/2}$ & $5.5\mu^{-1/2}$ & $5.0 a^{1/4} \mu^{-1/2}$ \\ 
\hline
$\Delta a_{\rm max}$ tadpole   & $\sqrt{8/3} \mu^{1/2}\simeq 1.63 \mu^{1/2}$ &$1.4\mu^{1/2}$ & $1.6\mu^{1/2}$ & $1.8\mu^{1/2}$ & $0.78 a^{3/4} \mu^{1/2}$ \\ 
\hline
$\Delta a_{\rm max}$ horseshoe & $2(3)^{1/6}\mu^{1/3}\simeq 2.40\mu^{1/3}$   &$4.6\mu^{1/2}$ & $4.7\mu^{1/2}$ & $5.0\mu^{1/2}$ & $1.8 a^{3/4} \mu^{1/2}$ \\ 
\hline
\end{tabular}
\end{small}
\caption{Comparison of properties of Lagrangian points $L_{1-5}$ and
orbits around them with those of their generalized versions
$L_{1-5}^N$. All quantities are given to lowest order in $\mu$. In
particular, expressions for the $N=2,3,4$ resonances were calculated
using a potential computed to first order in $\mu$ at $a=N^{2/3}$
rather than in the large-$a$ limit. The numerical values for the
generalized Lagrangian points and orbits are given for $C_J=3$.}
\label{reswidths}
\end{center}
\end{table}

\subsection{Generalized tadpoles}

The analogy is more obvious when motion around the fixed points is
investigated.  Small amplitude motion around the stable fixed point
$L^N_4$ and $L^N_5$ can be approximated by expanding $U^{1}$ around
its minimum.  This results in a harmonic oscillator equation:
\begin{equation}
\label{d2theta0dn2harmonic}
{d^2 \phi \over dn^2}
   = -6\pi a^{5/2}\mu 
     \left. \left( d^2U^1 \over d\phi^2 \right)\right|_ {\phi=\phi_{4,5}^N}
     (\phi-\phi_{\rm res}) 
\;\;\; .
\end{equation}
The small amplitude libration period around either $L^N_4$ or $L^N_5$
is therefore given by
\begin{equation}
\label{tlib}
K = {T_{\rm libration} \over 2\pi a^{3/2}}
  = \left( {3 \over 2 \pi} \left. d^2U^1 \over d\phi^2
                           \right|_{\phi=\phi_{4,5}^N} 
    \right)^{-1/2} a^{-5/4} \mu^{-1/2} 
  = 0.79 a^{-5/4} \mu^{-1/2}
\end{equation}
where in the last step we use $r_p=9/8$ in the large-$a$ limit to get
$d^2U^1/d\phi^2 \simeq 3.3$ at $\phi=\phi_{4,5}$. Note that $K$ gives
the number of periapse crossing per libration period. In our units,
where $2\pi$ is the period of the massive bodies, the libration period
is then $2\pi a^{3/2} K$.

Since Eq.~\refp{d2theta0dn2} describes motion under the influence
of a fixed potential, we can write down the conservation of energy
equation by multiplying Eq.~\refp{d2theta0dn2} by $d\phi\over dn$
and integrating with respect to $n$:
\begin{equation}
{1 \over 2} \left ({d \phi \over dn}\right ) ^2 
    + 6 \pi a^{5/2} \mu U^1 = {\rm constant} \;\;\; .
\label{pendenergy}
\end{equation}
The constant of integration is the `energy' associated with the
movement of the orbit in $\phi$ and $a$. Since the potential is
finite, it can only support a finite particle `speed' in libration
around $L^N_4$ or $L^N_5$. The `speed' is directly related to the
deviation of the semimajor axis from the resonance via
Eq.~\refp{dtheta0dn}, so the maximal width of these librations in $a$
is given by
\begin{equation}
\label{deltaamax}
\Delta a_{\rm max} 
  = \left( 4 \over 3\pi \right)^{1/2} \mu^{1/2} a^{3/4} 
    [U^1(\pi) - U^1(\phi_4)]^{1/2} 
  \simeq 0.78 a^{3/4} \mu^{1/2}  \;\;\; .
\end{equation}

\begin{figure}[!ht]
\begin{center}
\epsscale{.88}
\plotone{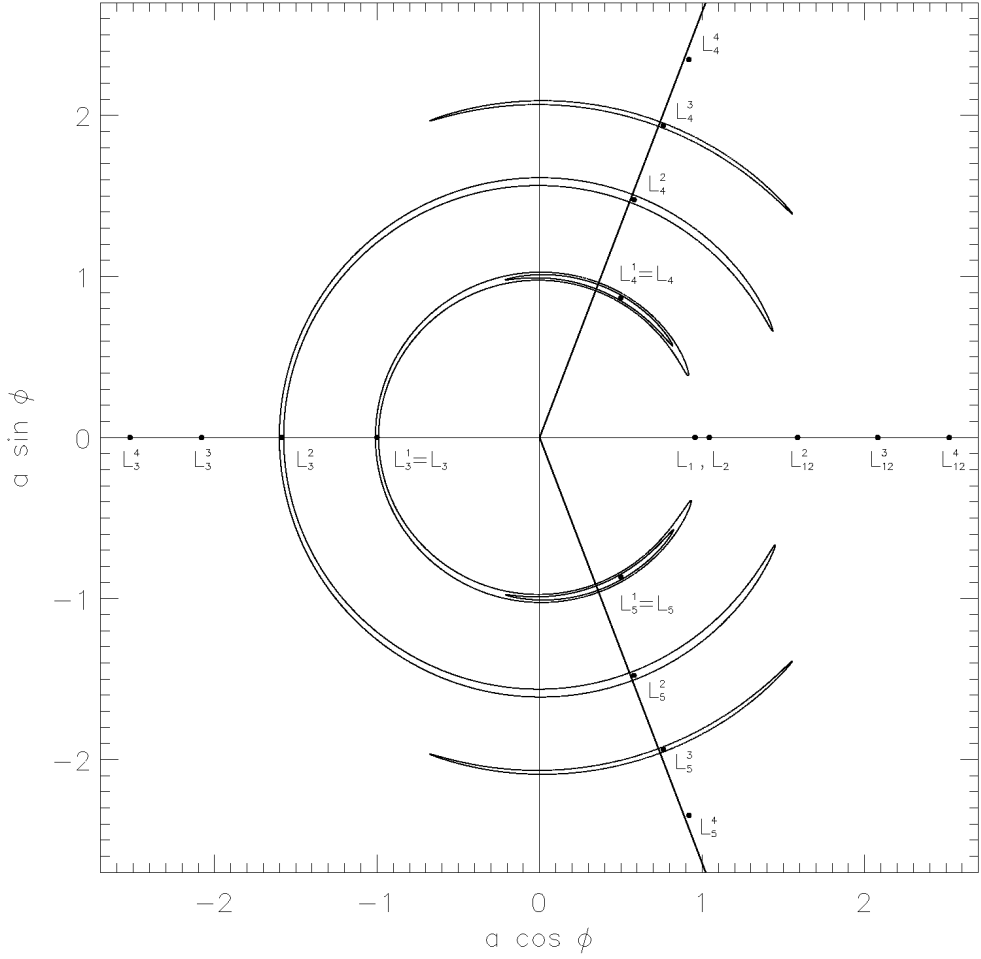}
\caption{The Lagrangian point analogues $L_i^N$ for $N=1,2,3,4$. The
diagonal lines trace the azimuths of $L_{4,5}^{\infty}$---that is, the
$\phi$ values of the minima in $U^1$.  Sample `horseshoes' and
`tadpoles' calculated via numerical integration with $\mu=2.5\times
10^{-4}$ and $C_J=3$ are also shown.}
\label{lagrangian_pts}
\end{center}
\end{figure}

These librations around the fixed points $L^N_4$ or $L^N_5$ are
analogues of the well-known tadpole orbits. Note that the maximal
widths of both the standard and generalized tadpole orbits scale as
$\mu^{1/2}$ (see Table~\ref{reswidths}). The similarity is more
apparent if we treat the $(a,\phi)$ parameters, which describe the
orbit of the particle, as polar coordinates as shown in
Figure~\ref{lagrangian_pts}. Seen in this way, $(a,\phi)$ are
analogous but not identical the polar coordinates of the particle in
the rotating frame: $a$ is the semimajor axis, not the radius, and
$\phi$ is the azimuth of the test particle in the rotating frame only
at periapse passage. Then the fundamental difference between the
$(a,\phi)$ plane and the rotating frame is that while generalized
Lagrangian points and the motion around them exist in a surface of
section made up of discrete points representing periapse passages, the
standard rotating frame with the standard Lagrangian points is made up
of continuous trajectories. Therefore, while the standard Lagrangian
points are fixed points in the rotating frame, the generalized points
represent periodic orbits in that frame.

The generalized tadpoles are equivalent to `asymmetric
librations'---trajectories whose resonant argument librates about a
value other than 0 or $\pi$. In this context, $L_4^N$ and $L_5^N$
correspond to `asymmetric periodic orbits' whose resonant argument is
constant but not equal to 0 or $\pi$. Our discussion above gives a
simple physical argument for the existence of asymmetric librations in
all stable $1:N$ exterior resonances. Again, note that the existence
of these asymmetric librations and asymmetric periodic orbits follows
from analysis of $U^1$ only when both the direct and indirect terms
are accounted for.

\subsection{Generalized horseshoes}

As the energy of the particle moving under the $U^1$ potential
increases beyond that of the maximal tadpole orbit, it overcomes the
lower potential barrier at $\phi=\pi$. As long as its energy is still
below the higher barrier at $\phi=0$, the particle will librate around
both the $L^N_{4}$ and $L_5^N$ points, avoiding only a narrow range in
$\phi$ around $\phi=0$. These trajectories are the generalized
horseshoe orbits. Using the same method as we used for the tadpoles,
we calculate their widths in the continuous approximation to be
\begin{equation}
\label{deltaamaxhorse}
\Delta a_{\rm max} 
   = \left (4\over 3\pi\right )^{1/2} \mu^{1/2} a^{3/4}  
     [U^1(0) - U^1(\phi_4)]^{1/2}
   = 1.8 a^{3/4} \mu^{1/2}
\;\;\; .
\end{equation}
The width of the maximal standard horseshoe does not follow this
$\mu^{1/2}$ pattern, since the standard horseshoe case differs
qualitatively from its generalized version. For the standard
horseshoe, the angular momentum change is concentrated near the
horseshoe's two ends. The close approach of the particle to the planet
there increases the strength of the interaction beyond $\mu$. As a
result, the width of the horseshoe scales as $\mu^{1/3}$ rather than
$\mu^{1/2}$. For a generalized horseshoe, the librating particle never
gets closer to the planet than $r_p-1$.

\begin{figure}[ht!]
\begin{center}
\epsscale{.95}
\plotone{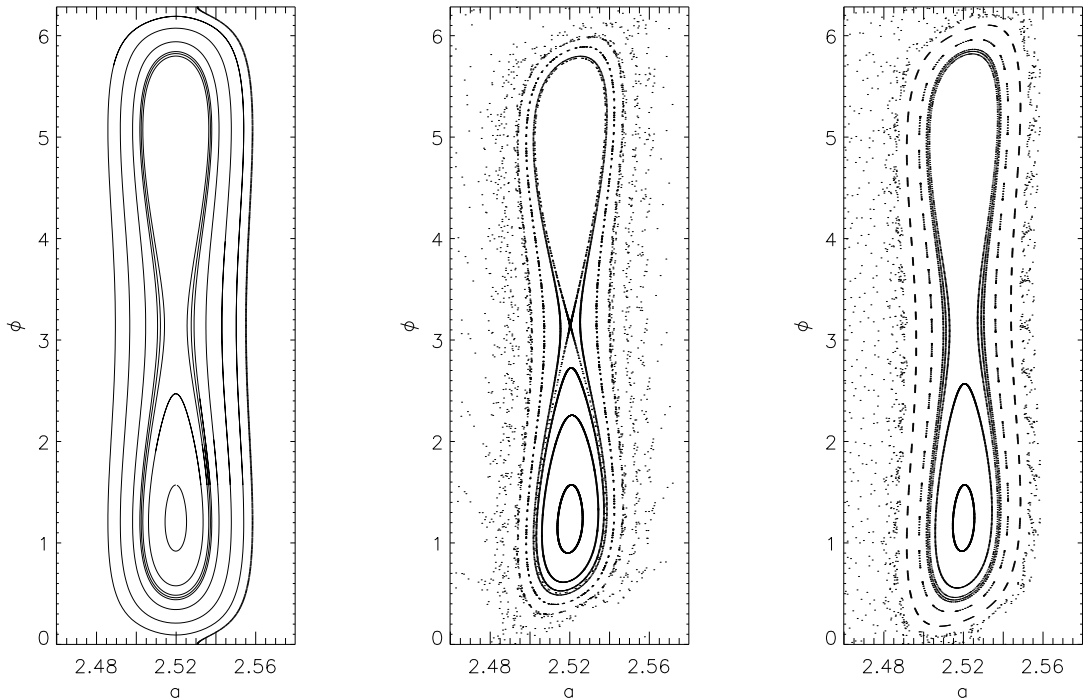}
\caption{$a$ vs. $\phi$ plot for $N=4$, $\mu=10^{-4}$, $C_J=3$. We use
$[0,2\pi]$ as the range in $\phi$ to show the trajectories more
clearly. The left-hand plot contains trajectories computed under the
continuous approximation. The middle plot contains a surface of
section computed via numerical integration of the orbits. The
right-hand plot contains trajectories computed via the eccentric
mapping discussed in \S\ref{mapping}. The same initial conditions were
used for the trajectories in all three plots. The continuous
approximation plot lacks the chaotic behavior evident in the numerical
integration and eccentric mapping plots. Trajectories in the mapping
plot differ from the numerical integration plot mostly because they
were calculated with $U^1$, the potential in the large-$a$ limit. 
%A similar mapping using the potential calculated to first order in
%$\mu$ at $a=4^{2/3}$ looks more similar to the numerical integration.
Note that the separatrix trajectory in the middle plot is chaotic but
on a scale too small to see in this figure (see Figure
\ref{athetasect2504_sep}).}
\label{athetasect_n4_mu0001}
\end{center}
\end{figure}

\begin{figure}[ht!]
\begin{center}
\epsscale{.95}
\plotone{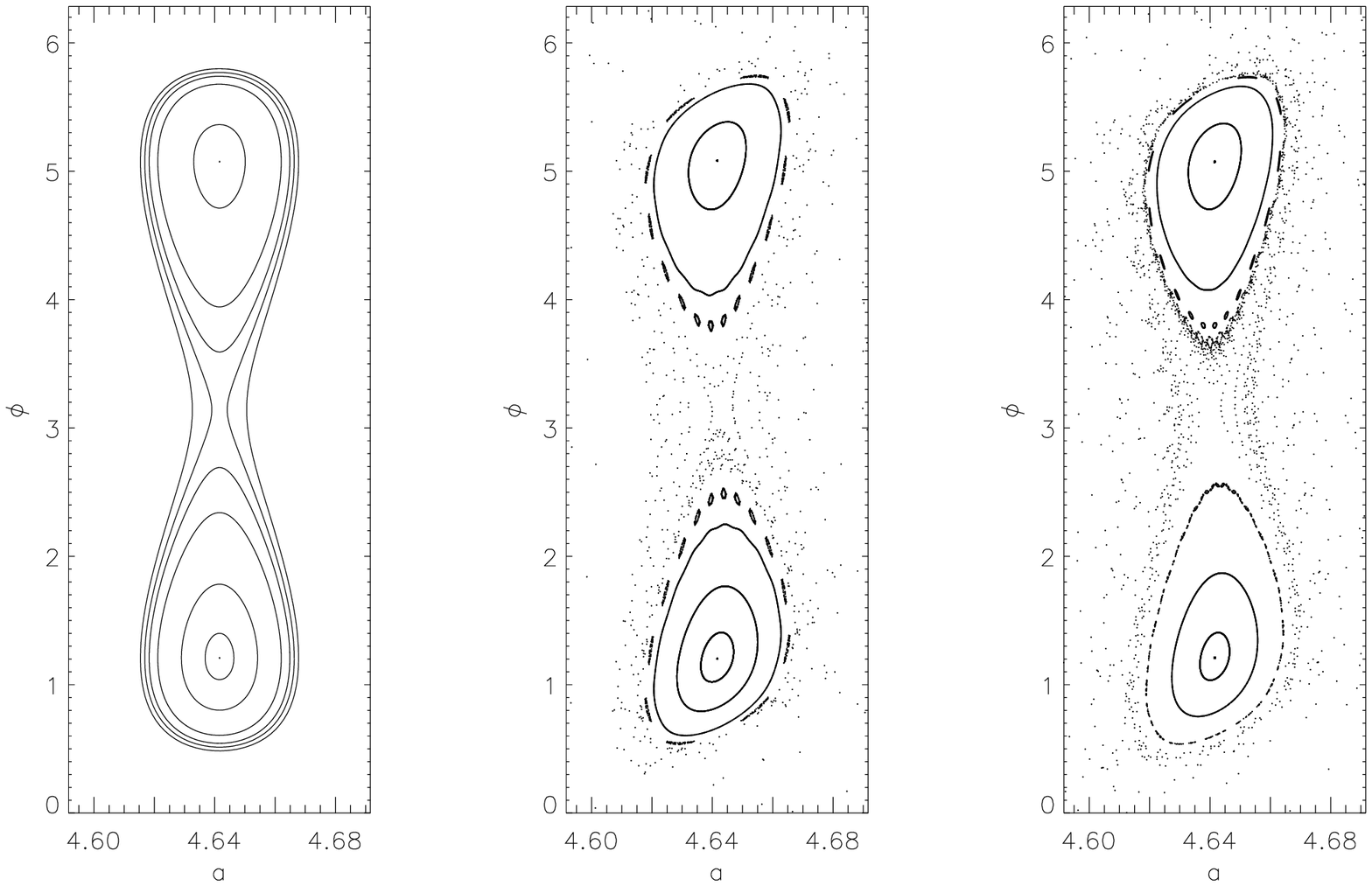}
\caption{Same as Figure \ref{athetasect_n4_mu0001} but for $N=10$. For
this larger $N$, the resonances are wider. So the resonance overlap is
more severe for the outer edges of the $N=10$ resonance than for the
$N=4$ one. This causes the destruction of all horseshoe orbits and the
distortion of the tadpoles relative to those computed in the
continuous approximation.}
\label{athetasect_n10_mu0001}
\end{center}
\end{figure}

In Figures~\ref{athetasect_n4_mu0001} and \ref{athetasect_n10_mu0001}
we show libration around $L^4_{4,5}$ and $L^{10}_{4,5}$. In order to
focus on the motion close to these points we plot $a$ and $\phi$ as
Cartesian rather than polar coordinates. In these plots, the
librations in the surfaces of section appear to be `warped' when
compared to the continuous approximations calculated using the
pendulum-like Eq.~\refp{d2theta0dn2}. This `warping' is due to the
discrete nature of the motion in the surfaces of section.  As a
trajectory moves from $\Delta a=0$ toward larger positive $\Delta a$
values, for example, the energy kicks stay positive and $\Delta a$
should keep increasing until the trajectory reaches a $\phi$ value
corresponding to a zero in the $\Delta E^1$ vs. $\phi$ curve. Within
the continuous approximation, we expect the trajectory to begin moving
back toward $\Delta a=0$ at exactly this $\phi$ because $\Delta E^1$
changes sign. A discrete trajectory will `overshoot' the nominal
$\phi$ where $\Delta E^1=0$ since a positive energy kick will carry
the trajectory past this $\phi$ before the first negative kick is
applied. As a result, the libration trajectories in the surfaces of
section tend to become warped in the direction in which orbits move
when librating. A quantitative discussion of this feature is
given in the next section.

%%%%%%%%%%%%%%%%%       4. ECCENTRIC MAPPING       %%%%%%%%%%%%%%%%%
\section{\label{mapping}The eccentric mapping}

The `warping' noted above suggests that the discrete nature of the
surface of section is essential to understanding some feature of the
motion in the $(a,\phi)$ plane. To study this, we define a mapping
from the $(a,\phi)$ plane to itself. Beginning at an arbitrary point,
this mapping produces an infinite series $(a^i,\phi^i)$ of points
visited by the test particle in the $(a,\phi)$ plane. Except perhaps
in the few lowest-$N$ resonances, we can build an excellent
approximation to the correct mapping by applying the first-order kicks
in the large-$a$ limit:
\begin{equation}
-{1 \over 2a^{i+1}}=-{1 \over 2a^i}+\mu \Delta E^1(\phi^i)
\end{equation}
\begin{equation}
\phi^{i+1}=\phi^i+2 \pi (a^{i+1})^{3/2}
\end{equation}
where the new value of $\phi$ is calculated modulo $2\pi$,
i.e. brought back into the interval $(-\pi,\pi)$ by adding an integer
multiple of $2\pi$. Note that the $a$-value used to find $\phi^{i+1}$
itself has index $i+1$; physically, this corresponds to the large-$a$
limit assumption that each energy kick is a discrete event associated
with a given periapse passage. Applying this mapping for several
initial values in the $(a,\phi)$ plane results in the right panels of
Figures~\ref{athetasect_n4_mu0001} and \ref{athetasect_n10_mu0001}.
The close resemblance between trajectories generated with the mapping
and with numerical orbit integrations demonstrate this mapping's
accuracy.

It turns out that the warping of the small amplitude tadpoles can be
understood completely in terms of the mapping. Close to the fixed
points $L^N_4$ and $L^N_5$, we define $\Delta a^i = a^i-a_{res}$
and $\Delta \phi^i=\phi^i-\theta_{res}$ so that the mapping becomes
\begin{equation}
\Delta a^{i+1}
 = \Delta a^{i} - 2 a_{\rm res}^2 \mu \Delta \phi^i
   \left. {d^2U^1 \over d\phi^2} \right| _{\phi=\phi^N_{4,5}}
\end{equation}
\begin{equation}
\Delta \phi^{i+1}=\Delta \phi^i + 3\pi a_{\rm res}^{1/2} \Delta a^{i+1}
\;\;\; .
\end{equation}
Since these are linear recursive equations, they can be solved
analytically by standard techniques. We seek a solution of the form
$(\Delta a^i,\Delta \phi^i)=(A,\Phi)\alpha^i$. (The reader should not
confuse indices with exponents---superscripts above Greek letters
other than $\phi$ are exponents.) Substituting in the recursive
equations, and seeking a non-trivial solution we obtain
\begin{equation}
\label{alpha}
(\alpha-1)^2 + 6\pi \mu a_{\rm res}^{5/2} 
               \left. {d^2U^1 \over d\phi^2} \right| _{\phi=\phi^N_{4,5}} 
               \alpha
= 0 \;\;\; .
\end{equation}
Note that the dimensionless parameter in this equation is simply
$(2\pi /K )^2$ where $K$ is the number of periapse passages per
libration in the continuous approximation as given by
Eq.~\refp{tlib}. If we denote the solutions as $\alpha_1$ and
$\alpha_2$, it is clear from the above equation that their product
$\alpha_1\alpha_2$ equals $1$. Since we are interested in potential
minima, $K^2>0$.

For $K\geq\pi$, the two roots are complex conjugates and each has
unity norm. The fixed point is therefore an elliptical point in the
discrete mapping as well as in the continuous approximation. The two
values of $\alpha$ are given by
\begin{equation}
\alpha_{1,2}={ 1- 2(\pi/K)^2 \pm 2  \sqrt{ 1-(\pi/K)^2} } (\pi/K) i \;\;\;.
\end{equation}
The number of periapse passages per libration is given by
\begin{equation}
\label{Kmap}
K^{\rm map} = {2\pi \over \arg (\alpha)}
  = 2\pi \left[ \arctan\left( {2  \sqrt{ 1-(\pi/K)^2}  (\pi/K) 
                               \over 1- 2(\pi/K)^2 }  \right) 
         \right]^{-1} \;\;\;.
\end{equation}
Clearly, as $K \rightarrow \infty$, $K^{\rm map}/K \rightarrow
1$. This is expected since the continuous approximation is justified
in this limit.  Using the two values of $\alpha$, we can find the
eigenvectors:
\begin{equation}
(A,\Phi)=\left( \mu a_{\rm res}^2 {d^2 U^1\over d\phi^2}, 
                (\pi/K)^2 \pm \sqrt{1-(\pi/K)^2}  (\pi/K) i
         \right )\;\;\; .
\end{equation}
Since the eigenvectors determine the axes of the ellipses representing
small librations about the fixed points, the similar shapes and
orientations of the smallest librations in the middle and right-hand
panels in each of Figures~\ref{athetasect_n4_mu0001} and
\ref{athetasect_n10_mu0001} confirm the eccentric mapping's accuracy.

For $a=10^{2/3}$, $\mu=10^{-4}$, the continuous approximation gives
$11.6$ orbits per tadpole libration (Eq.~\ref{tlib}). The eccentric
mapping gives $11.4$ orbits (Eq.~\refp{Kmap}). This is close to the
the 10.7 orbits per libration observed for very small librations about
the fixed points%
\footnote{The largest tadpole libration shown in
Figure~\ref{athetasect_n10_mu0001} breaks into 14 islands. This is an
example of the Poincar\'{e}-Birkhoff fixed-point theorem. It indicates
that this tadpole's libration period is 14 orbits. This lengthening of
the period is expected as the trajectory grows toward the separatrix
passing through $\phi=\pi$.}.
The negative power of $a$ in Eq.~\ref{tlib} implies that as $a$
increases, the number of periapse passages per tadpole libration
period will decrease and the trajectory shapes will become
increasingly warped.

In fact, when $a$ grows so large that $K$ falls below $\pi$, the
tadpoles are destroyed. For $K<\pi$, the roots of Eq.~\refp{alpha} are
real and distinct; therefore one of them is larger than unity. Then
the fixed point is not stable despite being at a potential
minimum. Our quantity $K$ is closely related to the residue $R$
discussed by \cite{jmg79}: $R=1-(\pi/K)^2$.

The warping of the tadpoles which, at its extreme, leads to
destruction of the resonances is absent in the continuous
approximation. However, it can be understood as perturbations from
nearby resonances. Interactions between neighboring first-order
resonances should become large enough to destroy these resonances when
the resonances begin to overlap. Eq.~\refp{deltaamaxhorse} implies
that as $a$ increases, the resonances widen in $a$ while the distance
between them decreases. Then we can find a condition on $\mu$ and $a$
for resonance overlap by dividing half the distance between
consecutive first order resonances by the width $\Delta a_{\rm max}$
of each resonance as given by equation \refp{deltaamaxhorse}.  In the
large-$a$ limit, the distance between resonances is given by ${2 \over
3} a^{-1/2}$ so we obtain
\begin{equation}
\label{1storderoverlap}
{{\rm resonance\; separation}\over 2\Delta a_{\rm max}}
  = \left (\pi \over 12 \right )^{1/2} 
    a^{-5/4} \mu^{-1/2} [U^1(0) - U^1(\phi_4)]^{-1/2}
  = 0.18 a^{-5/4} \mu^{-1/2} \;\;\; .
\end{equation}
This is proportional to $K^2$: in the large-$a$ limit with $r_p=9/8$,
the right-hand side is $0.28K^2$ and first-order resonances overlap
when $K<1.9$. In this case, therefore, first-order resonances are
destroyed before they formally overlap. With $\mu=10^{-3}$,
first-order resonances are destroyed above $a\simeq 4.02$. Then if we
neglect higher-order effects in $\mu$ and differences between the
potentials for $a\simeq 4$ and the large-$a$ limit, we expect just
eight stable first-order resonances.

%%%%%%%%%%%%%%%%%   5. HIGHER-ORDER RESONANCES   %%%%%%%%%%%%%%%%%
\section{\label{higher-order}Higher-order resonances}

As defined in \S\ref{DELTAE1}, higher-order resonances are the $p:p+q$
resonances with $p>1$. These resonances are located at $a_{\rm
res}=(N/p)^{2/3}$ where $N=p+q$ is an integer relatively prime to
$p$. In analogy to our treatment of first order resonances, we note
that if we neglect energy kicks, a particle exactly at resonance
should move in $\phi$ by $2\pi$ during each resonant cycle and by
$2\pi q/p$ between consecutive periapse passages. The stationary
points of this resonance should therefore occur at regular intervals
of $2\pi/p$ in $\phi$.

To study motion near but not at resonance, we include energy kicks.
For a particle close to resonance, we can follow its trajectory by
treating each resonant cycle as $p$ applications of the eccentric
mapping, one for each periapse passage in the cycle:
\begin{equation}
\label{deltaadeltaphip}
\left (
\begin{array}{c} \Delta a^{j+1} \\ \Delta \phi^{j+1}\end{array}
\right )
 =
\prod_{i=0}^{p-1}
\left (
\begin{array}{cc}
1                      & -2a_{\rm res}^2\mu 
                         \left. d^2 U^1\over d\phi^2 
                         \right |_{\phi = \phi_p^N+2\pi i/p} \\
3\pi a_{\rm res}^{1/2} & 1-6\pi a_{\rm res}^{5/2}\mu 
                         \left. d^2 U^1\over d\phi^2 
                         \right |_{\phi = \phi_p^N+2\pi i/p}
\end{array}
\right )
\left (
\begin{array}{c} \Delta a^{j} \\ \Delta \phi^{j}\end{array}
\right )
\;\;\; .
\end{equation}
As before, $\Delta a^j = a^j-a_{\rm res}$ and $\Delta\phi^j = \phi^j -
\phi_p^N$ where $\phi_p^N$ corresponds to the nearest fixed point in
the resonance. The condition under which the linearization in
$dU^1\over d\phi$ is valid is now $\Delta a^j\ll a^{1/2}/p^2$ instead
of $\Delta a^j\ll a^{1/2}$ because the number of energy kicks per
resonant cycle is $p$ instead of 1 and because the scale in $\phi$
over which the potential changes is now $\pi/p$ instead of $\pi$. The
condition under which linearization in $\mu$ is valid also changes
because the largest term linear in $\mu$ that appears in the mapping
matrix is $a_{\rm res}^{5/2}\mu {d^2 U^1\over d\phi^2}$. Though $\mu$
itself is small, cross-terms of order $\mu^2$ and higher are now
important unless $a_{\rm res}^{5/2}\mu {d^2 U^1\over d\phi^2}\ll
1$. This stronger condition is equivalent to $K\gg 1$, so the
higher-order resonance treatment does not offer any simplifying
advantages over the eccentric mapping discussed in \S\ref{mapping}
unless $K$ is large.

\begin{figure}[!ht]
\begin{center}
\epsscale{.75}
\plotone{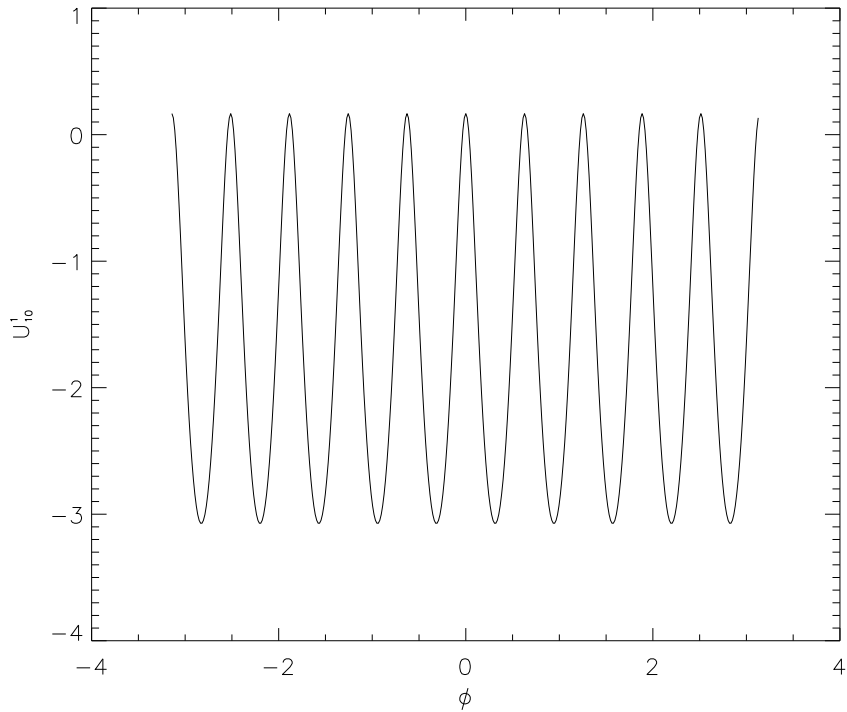}
\caption{$U^1_{10}$, or $U^1_p$ for a 10th-order resonance: $U^1$
summed over 10 consecutive periapse passages spaced evenly in
$\phi$. As before, $C_J=3$.}
\label{Un_mu001.10order}
\end{center}
\end{figure}

Since $\Delta\phi$ changes very little between consecutive periapse
passages in this $K\gg 1$ regime, we can use a variant of the
continuous approximation where we neglect the effects of drift in
$\phi$ within a single resonant cycle. Then we can treat the
particle's motion in terms of the net energy kick over an entire
resonant cycle rather than a single particle orbit. The net energy
kick is just the sum of $p$ energy kicks spaced $2\pi/p$ apart in
$\phi$, so the particle appears to move in the potential
\begin{equation}
U^1_p = \sum_{k=0}^{p-1} U^1(\phi-2\pi k/p) \;\;\; .
\end{equation}
Note that effects of the star's reflex motion do not contribute to
$U^1_p$ if $p>1$: the indirect term in $U^1$ is exactly sinusoidal and
the sum of $p$ identical sine curves spaced $2\pi/p$ apart in phase is
0, so $U^1_{p,{\rm ind}}=0$. Since the part of $U^1$ due to the
planet's direct contribution has just one maximum and one minimum at
$\phi=0,\pi$ respectively, $U^1_p$ has $p$ identical maxima and minima
(see Figure~\ref{Un_mu001.10order}. Then a trajectory librating in one
of the minima of $U^1_n$ should appear as a series of `islands' spaced
evenly in $\phi$ in the $(a,\phi)$ plane. As a result, no asymmetric
librations are possible in higher-order resonances. Our result that,
among exterior resonances, only $1:N$ resonances show asymmetric
librations is consistent with work done by \cite{cnf73}.  He analyzed
expressions for the time-averaged direct and indirect terms of the
disturbing function to find that asymmetric librations can exist only
in $p:p+q$ resonances where $p=\pm 1$.

\begin{figure}[!ht]
\begin{center}
\epsscale{.75}
\plotone{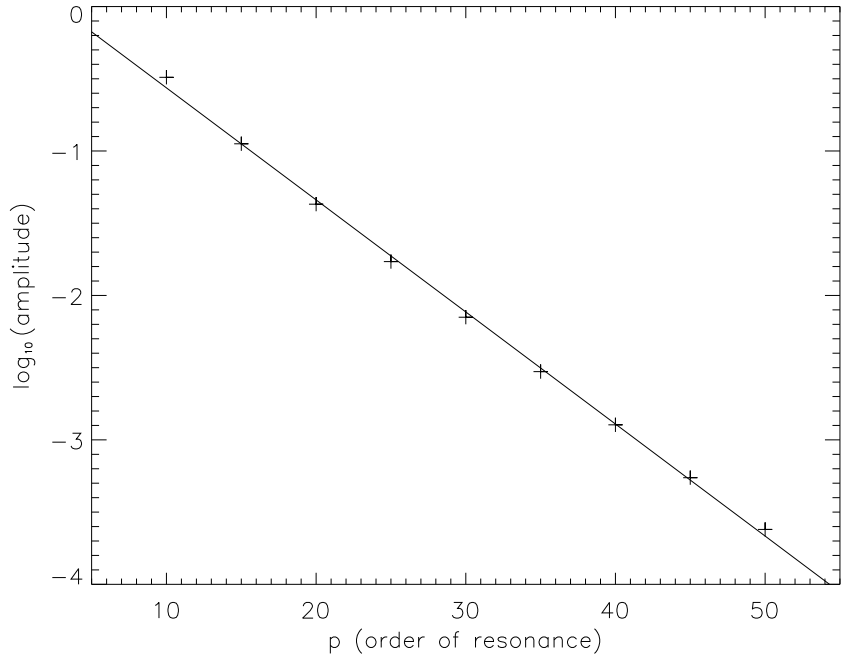}
\caption{Amplitudes of $U^1_p$ plotted on a log scale as a function of
$p$. Used $C_J=3$. Best-fit line is $\log_{10}[{\rm amplitude}] =
-0.078p+0.21$, or amplitude $\propto 1.2^{-p}$.}
\label{Un_amplitudes_mu001}
\end{center}
\end{figure}

Because the $p$ energy kicks received by the particle during each
resonant cycle are spaced evenly by $2\pi/p$ in $\phi$, we expect that
the kicks will partially cancel over each resonant cycle and that this
cancellation will improve exponentially as $p$ increases. We therefore
expect the amplitude of $U^1_p$ to decrease exponentially with $p$. As
Figure~\ref{Un_amplitudes_mu001} shows, this exponential decay is
observed numerically: a best-fit line in log-log space gives amplitude
$\propto 1.20^{-p}$.

%%%%%%%%%%%%%%%%%             6. CHAOS             %%%%%%%%%%%%%%%%%
\section{\label{CHAOS} Chaos in the large-\boldmath{$a$} limit}

We discuss just a few of the types and regions of chaos which arise
when $a$ is large. We first discuss `global' chaos, which consists of
chaotic regions that span a few resonance widths or more. We then give
a few examples of `local' chaos---chaos confined to regions within a
single resonance---and compare the structure seen in trajectories in
the $(a,\phi)$ plane on local and global scales.

On large scales in $a$, chaotic regions arise where there is overlap
between neighboring resonances or instability due to a small winding
number as discussed in \S\ref{mapping}. In regions of the $(a,\phi)$
plane where first-order and/or higher-order resonances overlap even
partially, we expect to see contiguous `globally' chaotic regions
which span large ranges in $a$. Any remaining stable regions within
resonances will appear as `islands' of stable librations. Particles
can undergo large changes in $a$ only if they move in these chaotic
regions, so such regions provide the only channels through which
initially bound particles can escape from the star-planet system.

If $a$ is large enough, $K$ falls below $\pi$ and these `islands'
disappear as discussed in \S\ref{mapping}. For a given value of
$\mu$, we see from Eq.~\ref{tlib} that
this occurs when
\begin{equation}
\label{res_unstable}
a > \mu^{-2/5} \left ({3\pi\over 2}{d^2 U^1\over d\phi^2}\right )^{-2/5}
  = 0.33 \mu^{-2/5}
\end{equation}
where, again, the numerical example corresponds to $r_p=9/8$. The
condition $K<\pi$ for resonance destruction in higher-order resonances
does not follow simply from Eq.~\ref{1storderoverlap}: effects of
order $\mu^2$ or higher may make the winding number expressions for
higher-order resonances differ from the first-order resonance case in
Eq.~\ref{tlib}. However, numerical experiments suggest that the $a$ at
which higher-order resonances become unstable is comparable to but less
than that given by \ref{res_unstable}.

\begin{figure}[ht!]
\begin{center}
\epsscale{.75}
\plotone{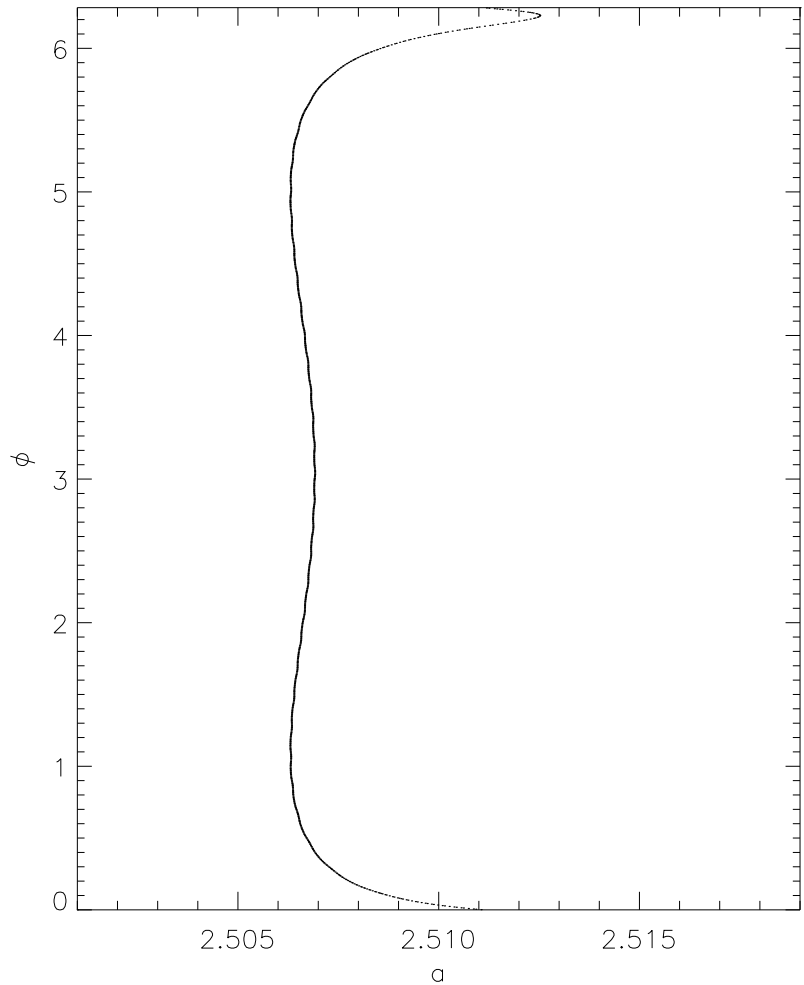}
\caption{A stable circulating trajectory in $(a,\phi)$ space
calculated via numerical integration with $\mu=5\times 10^{-6}$ and
$C_J=3$.}
\label{stablecirc_mu5e-6}
\end{center}
\end{figure}

If $\mu$ is small enough, there should be regions in $a$ where the
resonances do not overlap. In these regions we expect to see stable
trajectories which circulate around the resonances instead of
librating in them. We find numerically that stable circulating
trajectories exist for $\mu$ values up to at least $\mu=5\times
10^{-6}$; an example is shown in Figure~\ref{stablecirc_mu5e-6}.
\cite{jmg79} suggests that as $\mu$ increases, the last stable
circulating trajectory should have semimajor axis $a$ such that
$a^{3/2}$ is the golden ratio $(1+\sqrt{5})/2$. Our situation differs
qualitatively from Greene's in that our potential depends on its
linear coordinate, the semimajor axis, while Greene's potential, which
is given by the standard map, is independent of its linear coordinate
$r$. Specifically, when $a$ is not much larger than 1, $r_p-1 \ll 1$;
this leads to a larger maximum energy kick and potential well depth
than is expected for $r_p=9/8$, so the resonances are wider and more
prone to overlap for a given $a$ close to 1 than we would expect in
the large-$a$ limit. However, as $a$ increases the resonance spacing
decreases as discussed in \S\ref{mapping}. These competing effects
suggest that the last stable circulating trajectories---those which,
in a sense, are `farthest' from any resonances---should lie neither
near $a=1$ nor at $a\gg 1$. Also, effects of order $\mu^2$ and higher
which are present in our situation have no analogue in Greene's
analysis of the standard map. So it is unsurprising that the last
stable circulating trajectories which we found numerically have
$a^{3/2}$ unrelated to $(1+\sqrt{5})/2$.

Continuity and uniqueness imply that in a system with two degrees of
freedom, stable trajectories in a two-dimensional surface of section
cannot be crossed. In the planar restricted three-body problem,
therefore, stable circulating trajectories divide the $(a,\phi)$ plane
into separated regions in $a$. This implies that for any $\mu<5\times
10^{-6}$, chaotic and regular trajectories which start close enough to
the planet are confined to a set range in $a$. Then the particles
associated with these trajectories can never escape from the
star-planet system.

\begin{figure}[ht!]
\begin{center}
\epsscale{.75}
\plotone{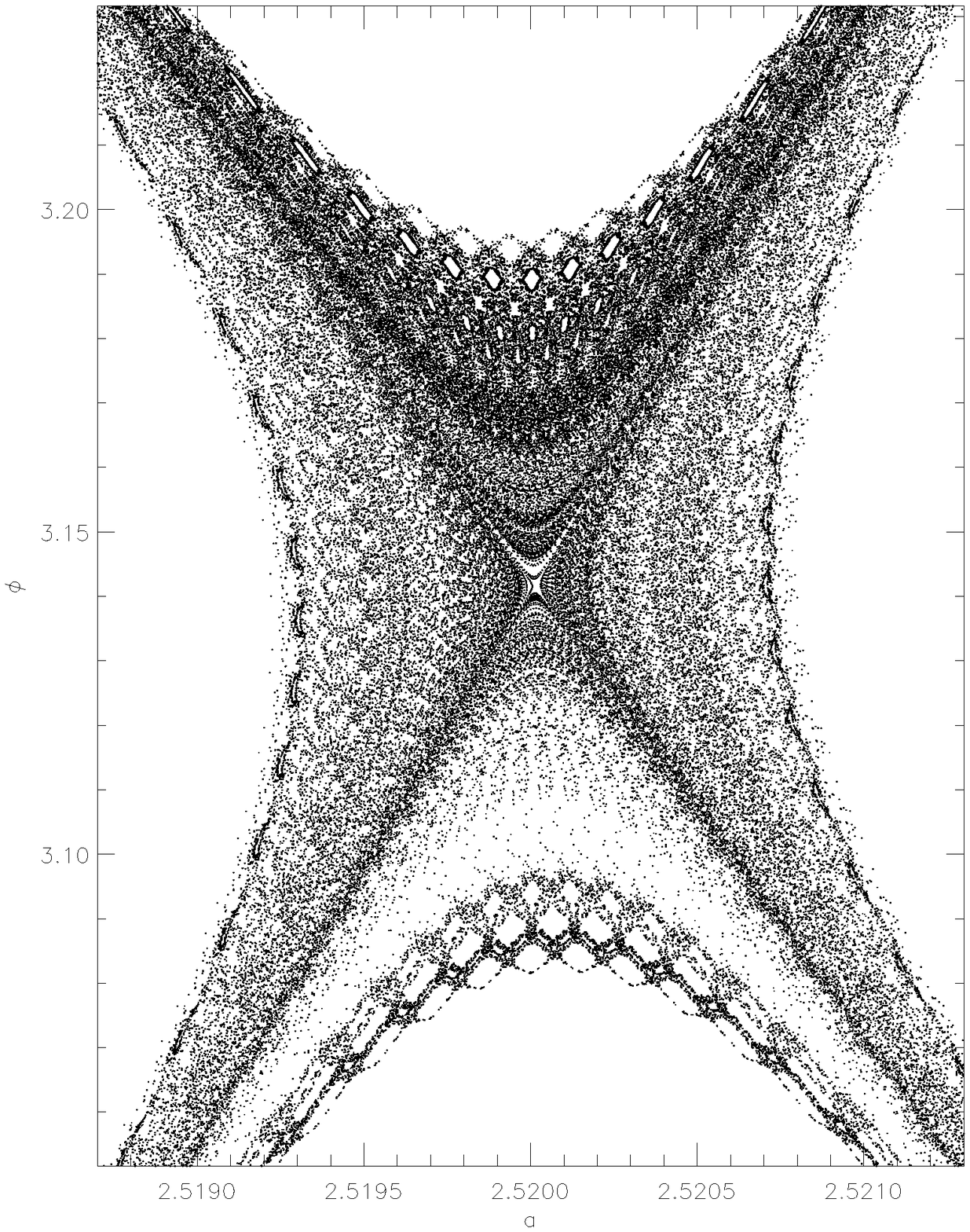}
\caption{A single chaotic trajectory corresponding to the separatrix
dividing `tadpole' and `horseshoe' librations in the $N=4$ resonance
when $\mu=10^{-4}$ and $C_J=3$. This trajectory was computed via
numerical integration with the same initial conditions as were used to
produce the separatrix trajectory in the middle panel of
Figure~\ref{athetasect_n4_mu0001}. It is confined in the $(a,\phi)$
plane by stable librations similar to the smallest horseshoe and
largest tadpoles shown in Figure~\ref{athetasect_n4_mu0001}. Note the
empty spots in the outer reaches of the chaotic trajectory; these
`avoided' areas correspond to islands of stable librations around
stable fixed points in trajectories with rational winding number.}
\label{athetasect2504_sep}
\end{center}
\end{figure}

This bounding of chaotic regions by stable trajectories also leads to
confinement of chaos on very small scales in $a$. Regions of
small-scale `local' chaos arise from unstable fixed points which must
be saddle points due to the area-preserving nature of the eccentric
mapping; the separatrices associated with the saddle points are
chaotic. If stable continuous trajectories exist near a saddle point,
they act as boundaries to the chaotic separatrix. Prominent examples
of these separatrices include those dividing tadpole and horseshoe
analogue trajectories within individual first-order resonances. These
regions are bounded by the largest stable tadpole and smallest stable
horseshoe, so their maximum range in $a$ is at most the resonance
width. One of these is shown in Figure~\ref{athetasect2504_sep}.

The existence of similar separatrices on all scales smaller than a
single resonance width follows from the Poincar\'{e}-Birkhoff theorem,
which states that for small enough $\mu$, a trajectory with rational
winding number $K$ is associated with equal numbers of alternating
stable and unstable fixed points. According to the KAM theorem, some
continuous trajectories---that is, trajectories with irrational
$K$---should also be stable as long as $\mu$ is small enough. If a
trajectory with rational $K$ is bounded on either side by stable
continuous trajectories with irrational $K$, then the chaos associated
with the unstable fixed points is confined to the region bounded by
the continuous trajectories. As for the stable fixed points, they are
associated with their own librating trajectories; the tadpole analogue
made up of islands shown in Figure~\ref{athetasect_n10_mu0001} gives
an example of such librations. We expect some librations like these to
have rational winding numbers and, therefore, their own sets of
unstable fixed points and confined chaos on an even smaller scale. In
principle, this argument can be applied repeatedly within a single
resonance to unearth similar chaotic regions on scales as small as
desired.

We can treat the entire $(a,\phi)$ plane as an extension of this
self-similarity to the largest possible scales.  If we plot $(a,\phi)$
as polar coordinates, a $p:p+q$ resonance trajectory appears to `wind'
around the point $a=0$ with 
%period $2\pi(p+q)/p$---that is, the orbital period of the resonance. 
%This period corresponds to a 
rational winding number $p/(p+q)$. Also, the corresponding resonance
is associated with $p$ stable and $p$ unstable fixed points when $p>1$
and $2$ stable and $2$ unstable fixed points when $p=1$. This provides
a striking visual analogy to the librations seen within a single
resonance.

%In other words, the $(a,\phi)$ plane itself appears like a set of
%librations about a fixed point when $a=0$ is associated with the
%central `fixed point' and $\phi$ is the angular variable. The one
%irregularity occurs for $p=1$ resonance trajectories, which have twice
%as many stable and unstable fixed points as we would expect from the
%$p>1$ trajectories.

%%%%%%%%%%%%%%%%%           7. DISCUSSION          %%%%%%%%%%%%%%%%%
\section{\label{DISCUSSION}Discussion and conclusions}

Using simple physical reasoning instead of explicit analysis of terms
in the disturbing function, we have developed a framework for studying
particles perturbed into exterior high-eccentricity orbits in the
circular planar restricted three-body problem. We have found that, to
first order in $\mu$, these orbits move in $(a,\phi)$ phase space
according to a potential with maxima at $\phi=0,\pi$ separated by
symmetrical minima. In the special case of resonance orbits, movement
in this potential translates into behavior governed by a modified
pendulum equation. Previous pendulum-analogue analyses of this problem
have usually been formulated via the disturbing function and the
continuous resonant argument \citep{ocw97a,sfd83}.

Our analysis, specifically that of mapping, is most similar to that of
\cite{lm99}. They consider the evolution of high-eccentricity
comet-like orbits in the low-inclination circular restricted
three-body problem by integrating numerically to find the energy kick
as a function of the resonant angle at periapse and then using this
energy kick to create a mapping which takes one periapse passage to
the next. However, \cite{lm99} are interested in particle orbits that
cross the orbit of the secondary, so the form of their energy kick is
qualitatively different from ours. In particular, while a small
nonzero orbital inclination would barely affect our energy kick
function, it could drastically change the shape of the overall energy
kick function in the case of planet-crossing orbits. Partly because of
this, they do not discuss their energy kick it in terms of a
potential. Also, they focus on the chaotic diffusion of the particle
toward escape or capture rather than on motion in resonances.

For $1:N$ resonance orbits---that is, those we call first-order
resonance orbits---the shape of the potential generates analogues of
the Lagrangian points for $N>1$ resonances. The potential similarly
leads to two kinds of libration analogous to the horseshoe and tadpole
orbits seen in a $1:1$ resonance. $p:N$ resonances---that is, those we
call higher-order resonances---show only one kind of libration: when
the winding number is large, the sum relating the higher-order
resonance potentials to the first-order resonance potential eliminates
the indirect term responsible for the tadpole analogues. 

Several authors discuss the tadpole analogues' presence or absence in
mean-motion resonances in general under the name `asymmetric
librations'; \cite{dn01} are the only others we know of to refer to
the $1:N$ resonance librations as `tadpoles' and `horseshoes', though
they do not elaborate on this analogy. Some authors have used
analytical studies of the Hamiltonian and the disturbing function to
set conditions for the existence of asymmetric resonances
\citep{adb94,cnf73,pjm70}. In particular, \citet{cnf73} analyzed the
time-averaged direct and indirect parts of the disturbing function to
deduce that only what we call first-order resonances should show
asymmetric librations. \cite{adb94} also found analytically that
asymmetric librations only exist in what we call first-order exterior
resonances. We confirm this and provide a simple physical explanation.

Others have used numerical methods to confirm the existence of
asymmetric librations for particular $1:N$ resonances and ranges in
eccentricity \citep[see, for
example,][]{ocw97b,cb94,pjm78,cnf73,pjm58}. Some of these also
compare their numerical results to expressions for the Hamiltonian
correct to first or second order in eccentricity. Although the
agreement is generally good for what we call first-order resonances,
the Hamiltonian expressions for what we call higher-order resonances
tend to predict spurious asymmetric librations.  We believe these are
due to extra extrema introduced into the potential by the $\cos
(2\cdot{\rm resonant\; angle})$ term in the Hamiltonian.  In some of
the more recent studies involving asymmetric librations
\citep{dn01,rm96} the discussion is framed in terms of the
dynamics of the classical Kuiper Belt and so is confined mostly to
what we call low-$N$ first-order and low-$p$ higher-order resonances
in the low- to moderate-eccentricity regime.

We find a limit on $a$ for stable first-order resonances. Overlap
between the resonances creates chaotic regions of $(a,\phi)$ phase
space; for semimajor axes larger than some $a\propto \mu^{-2/5}$, the
resonance centers are overlapped and no stable librations are
possible. This is the high-eccentricity analogue of the well-known
chaotic criterion $|a-1|\simeq \mu^{2/7}$ found by \cite{jw80} for the
circular planar restricted three-body problem in the low eccentricity
case. We use the Chirikoff criterion for resonance overlap to estimate
the location of the onset of chaos. For sufficiently narrow
resonances, or small enough $\mu$, there exist regions in $(a,\phi)$
space which lie outside all of the resonances but which are not
chaotic. In the planar problem we consider, particles interior to the
circulating trajectories in these regions are never able to escape
from the star-planet system.

The basic framework for the behavior of high-eccentricity orbits and
the properties of chaotic regions in $(a,\phi)$ space can be applied
to the orbital evolution of small bodies in the solar system. Objects
in the Kuiper Belt, for example, are believed to have arrived there
via interactions with Neptune \citep[see, for example,][]{rm00}; we
can apply this framework to study their trajectories. Many of these
objects are known to be in resonances \citep[see][for a recent
compilation]{eic03}. The mass ratio between Neptune and the Sun is
$\mu_N=4.4 \times 10^{-5}$. Since this is above the critical $\mu
\approx 5 \times 10^{-6}$, Kuiper Belt objects with $C_J=3$ are either
librating around a resonance or moving chaotically. The latter could,
in principle, be ejected as there is no stable circulation for that
value of $\mu=\mu_N$ and $C_J=3$. However, the known Kuiper Belt
objects span a range in $C_J$ roughly $2.6<C_J<3.2$. In the planar
problem, $\mu=\mu_N$ and for example $C_J=3.1$, stable circulations
exist, and protect some of these objects from escape. To study the
ultimate fate of such Kuiper Belt objects, the effect of inclination
must be understood.

Similarly, we might apply this framework to the scattering of small
planetesimals by giant protoplanets and could provide insight to
numerical integrations such as those of \cite{far96,ebf01}. Studies
like this require an investigation of the way in which the energy
kicks move orbits through the `global chaos' region surrounding the
resonances in the $(a,\phi)$ plane. Although the antisymmetry of
$\Delta E^1(\phi)$ about $\phi=0$ suggests that the orbits should
random walk through phase space, effects of nearby resonances
\cite[e.g.][]{lm99} and terms of higher order in $\mu$ become
important on timescales long enough for escape to become possible. The
importance of second order effects may be understood as follows. Since
the amount of extra energy needed to escape is $1/a_{\rm init}\sim 1$
and the energy kick per orbit is $\sim\mu$, we expect that the average
number of kicks needed to escape is $\sim\mu^{-2}$. Note that unlike
the first order kicks, the $\mathcal{O}(\mu^2)$ energy kicks do not
average to 0 over the interval $(-\pi,\pi]$ in $\phi$. This is also
apparent from Figure~\ref{deltaE_mu001}. Therefore, with $\mu^{-2}$
kicks, the sum of $\mathcal{O}(\mu^2)$ effects produced by the energy
kicks will be of order unity---that is, of size comparable to the
total first order effect.

\acknowledgments MP is supported by an NSF Graduate Research
Fellowship and RS holds a Sherma Fairchild senior research
fellowship. We thank Peter Goldreich for useful discussions.

\bibliographystyle{apj}
\bibliography{resorbit}

%%%%%%%%%%%%%%%%%            LOOSE ENDS            %%%%%%%%%%%%%%%%%
\begin{large}
\end{large}

\end{document}